\documentstyle[]{aa}
\input psfig.sty

\begin{document}
%\setcounter{page}{01}

%\thesaurus{06     % A&A Section 6: Form. struct. and evolut. of stars
%              (08.22.1;  % Stars: variables: Cepheids
%               08.15.1;  % Stars: oscillations
%               08.04.1)}  % Stars: distances

%
\title{H$_2$O maser emission from bright rimmed clouds in the northern
hemisphere\thanks{Based on observations obtained with the 32-m Medicina
radiotelescope.}}

%   \subtitle{.........}

\author{R. Valdettaro\inst{1} \and F. Palla\inst{1} \and 
J. Brand\inst{2} \and R. Cesaroni\inst{1} }

\institute{INAF-Osservatorio Astrofisico di Arcetri, Largo E. Fermi, 5, 
I-50125 Firenze, Italy 
\and
INAF-Istituto di Radioastronomia, Via Gobetti 101, I-40129 Bologna, Italy
}

\offprints{R. Valdettaro, \email{rv@arcetri.astro.it}}

\date{Received/Accepted}

\abstract{
We report the results of a multi-epoch survey of water maser observations at
22.2 GHz with the Medicina radiotelescope from 44 bright rimmed 
clouds (BRCs) of
the northern hemisphere identified by Sugitani et al. (1989) as potential
sites of star formation.  The data span 16 years of observations and allow to
draw conclusions about the maser detection rate in this class of objects.  In
spite of the relatively high far-infrared luminosities of the embedded
sources ($L_{\rm FIR}\ga 10^2$~L$_\odot$), H$_2$O maser emission was detected
towards three globules only.  Since the occurrence of water masers is higher
towards bright IRAS sources, the lack of frequent H$_2$O maser
emission is somewhat surprising if the suggestion of induced intermediate-
and high-mass star formation within these globules is correct.  The maser
properties of two BRCs are characteristic of exciting sources of low-mass,
while the last one (BRC~38) is consistent with an intermediate-mass object.
We argue that most BRCs host young stellar objects of low-luminosity, likely
in an evolutionary phase later than the protostellar Class 0 sources, 
and that a significant
contribution to the observd IRAS luminosity comes from warm dust heated by
the radiation from the bright rim.
}

\titlerunning{H$_2$O maser emission from BRCs}
\authorrunning{Valdettaro et al.}

\maketitle

%
%________________________________________________________________

\section{Introduction}

Bright rimmed clouds (BRCs) have been considered very promising sites of star
formation, perhaps induced by the compression of ionization and shock fronts
from nearby H{\sc ii} regions (Bertoldi 1989). The boundary layer between the
neutral gas and the gas ionized by the incident photons is often called {\it
bright rim}, but the clumps are also classified as cometary globules or
elephant trunks depending on their appearance (Tauber et al. 1993).  The
ionization front responsible for the appearance of the bright rim, also
induces gas heating, collection and compression that may ultimately lead to
triggered gravitational collapse (Elmegreen 1998, Vanalha \& Cameron 1998).

The outcome of this process critically depends on the balance between the
pressure due to the ionization front and the internal pressure due to
turbulent motions.  An analysis of the internal and external pressures on a
group of BRCs indicates that the conditions of approximate pressure
equilibrium are prevalent, with photoionization shocks likely propagating
inside the clouds (Morgan et al. 2004).  However, direct
evidence of infalling gas in BRCs is still lacking.  The results of a
multiline molecular line survey of BRCs by De Vries et al. (2002) indicate a
very low percentage of clouds with the blue-asymmetric profile in optically
thick lines due to inward motions.  Therefore, it is still uncertain whether
star formation occurs in BRCs in response to an external trigger or in a
spontaneous mode, as in more quiescent dark clouds.  
%De Vries et al. suggest that the lack of the infalling signature may be 
%due to the additional heating from the nearby H{\sc ii} region.

Observationally, BRCs are sites of ongoing star formation as, in
addition to embedded IRAS sources, molecular outflows and Herbig-Haro objects
(Sugitani, Fukui \& Ogura 1991, hereafter SFO;
Ogura et al. 2002), they frequently contain small clusters of near-IR
stars (Sugitani et al.  1995; Thompson et al. 2004).  There is also
evidence that the
embedded IRAS sources have intermediate- to high- far-infrared (FIR)
luminosities, $L_{\rm FIR}\ga 10^2$ L$_\odot$ (SFO). Finally, the ratios of
the luminosity of the IRAS sources to the globule mass (as measured from
molecular line emission) appear to be much higher than those found in
isolated dark globules (Sugitani et al. 1989).  Altogether, these properties
suggest that star formation may proceed in a different mode in globules
associated with bright rims than in more quiescent ones, which spawn
low-luminosity objects ($L_{\rm FIR}\la 10^2$ L$_\odot$).  Our aim is to test
this hypothesis by using water masers as a reliable diagnostic of the
presence and nature of newly formed objects.

Water maser emission is associated with the early stages of star formation of
YSOs of all luminosities (e.g. Palagi et al. 1993; Furuya et al. 2001).
Long-term monitoring of H$_2$O masers in SFRs (Valdettaro et al. 2002; Brand
et al. 2003) has shown that for FIR luminosities of the magnitude found in
the BRCs (at least for $L_{\rm FIR}\ga 10^2$ L$_\odot$) the detection rate of
water masers is about 75--80\%, and drops substantially at lower
luminosities. In addition, in the latter sources the frequency of occurrence
of maser emission is strongly episodic (Claussen et al. 1996).  For example,
the high sensitivity (rms$\sim$0.23 Jy) search of Furuya et al. (2003) using
the Nobeyama 45-m telescope detected H$_2$O maser emission towards 40\% of
their sample of Class 0 sources, but only 4\% for Class I and none for Class
II sources.

For our study, we have observed at least three times all of the 44 BRCs
listed by SFO, visible in the northern hemisphere ($\delta>-30^{\circ}$).
The list of BRCs with the observing dates and the individual r.m.s. are given
in Table~1. We detected water maser emission in three sources, two of which
(BRC~30 and BRC~36) represent first time detection.

\section{Observations}
Observations of the maser emission from the $6_{16}-5_{23}$ rotational
transition of water at 22 GHz were performed with the Medicina 32-m antenna
\footnote{The 32-m VLBI antenna at
Medicina is operated by the INAF-Istituto di Radioastronomia in Bologna.}
in many sessions between March 1989 and June 2005.  
During this 16 year interval, all sources have been observed at least three
times.  Spectra were taken in position-switching mode with typical integration
times of 5 min both on- and off-source. Starting in 2001, in order to improve
the S/N of the observations, the integration time on each source was
increased to 20 min (both on- and off-source).

\noindent
The telescope HPBW at 22.2~GHz is 1.9$^{\prime}$.  The pointing accuracy is
better than $\sim 25^{\prime\prime}$. Flux calibration was derived from
antenna temperature measurements at different elevations of the continuum
source DR21, whose flux density is assumed to be 18.8 Jy (Dent 1972). The
calibration uncertainty is estimated to be $\sim 20$\%. For details of the
observational parameters, we refer to Valdettaro et al. (2002).

\begin{figure}
\psfig{file=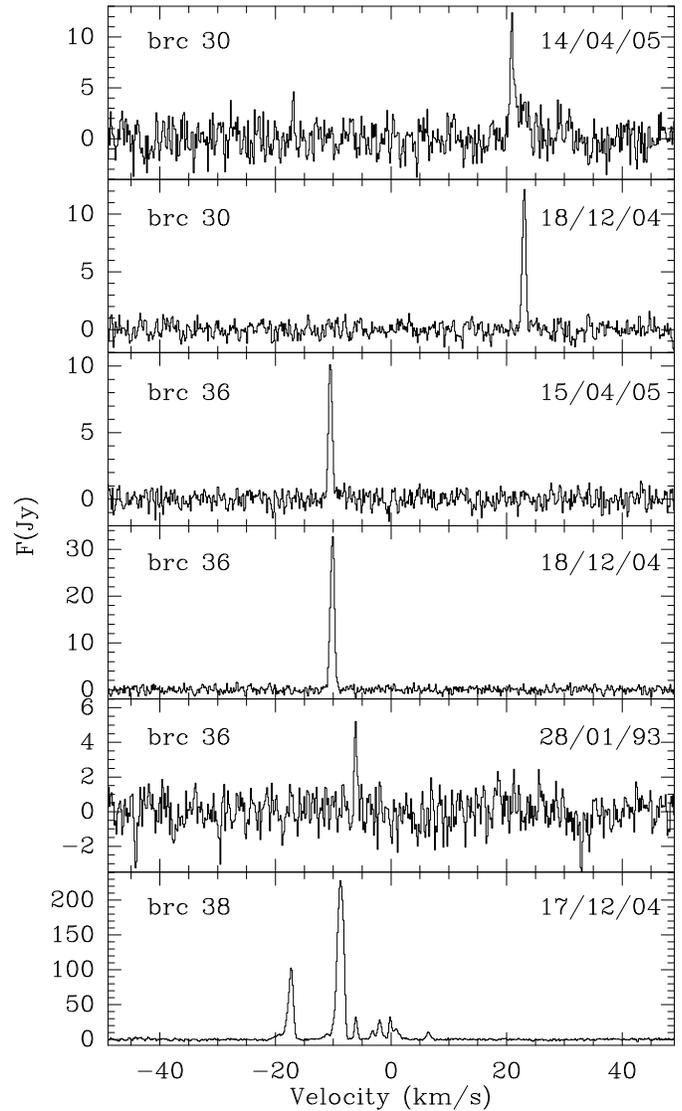,width=88mm}
\caption{Detected H$_2$O masers in BRCs. The observation date is indicated
in each panel. \label{det}}
\end{figure}

\begin{figure}
\psfig{file=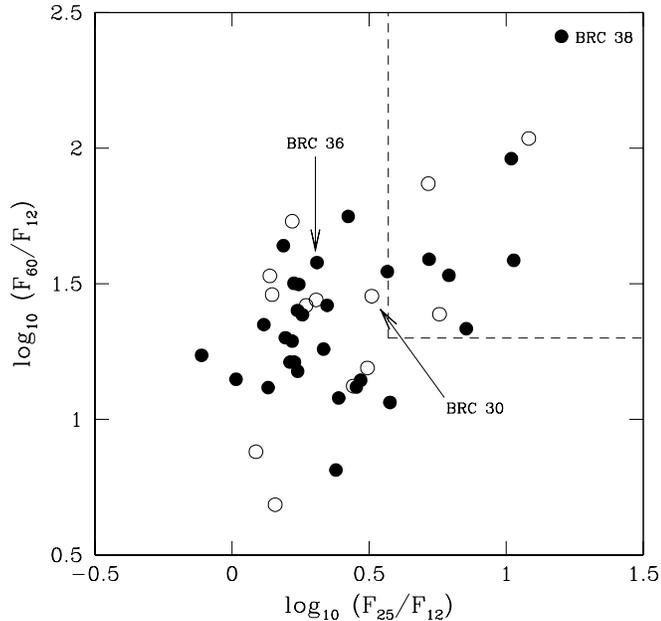,width=88mm}
\caption{Location of the 44 BRCs in the IRAS color-color
diagram. Solid dots represent sources with good IRAS fluxes, while
those with upper limits on at least one of the IRAS fluxes are shown as 
empty circles. The dashed lines delimit
the boundaries of UCH{\sc ii} regions proposed by Wood \& Churchwell (1989).
The three H$_2$O maser sources are labeled. \label{colors}}
\end{figure}

\section{Results}
Water maser emission is detected towards Sugitani's clouds
BRC~30, BRC~36, and BRC~38.  Their spectra are shown in Figure~1 and the main
parameters of the maser emission are listed in Table~2 which gives the BRC
number (Col. 1), the observing date (Col. 2), the spectral resolution
(Col. 3), the $1\sigma$ r.m.s. (Col. 4), the minimum and maximum of the 
velocity interval where emission is observed (Cols. 5-6), the velocity of
the peak flux (Col. 7), the total integrated H$_2$O flux density (Col.8),
and the maser luminosity (Col. 9). While H$_2$O maser
emission was detected only once in BRC~30 and three times in BRC~36, 
the maser in BRC~38 has been active at all times since the beginning of the
observing campaign. In the case of BRC~30 and 36 the emission displayed in
Fig.~1 represents a first time detection, while the maser in BRC~38 was 
first found by Felli et al. (1992). We note that 5 BRCs were also
included in Furuya et al. (2003) survey towards low-mass YSOs (BRC 16, 18, 38,
39, 44). These objects were repeatedly observed by Furuya et al. in 1998 and
H$_2$O maser emission was detected only in the case of BRC~38.

Below, we briefly discuss the main features of the masers associated with
BRC~30 and BRC~36, while BRC~38 will be the subject of a separate analysis
(Valdettaro et al. 2005b).  The distribution of the 44 BRCs in the FIR
color-color diagram is displayed in Figure~2, and those with H$_2$O maser
emission indicated as labeled.  The properties of the BRCs and their
embedded IRAS sources are listed in Table~3.

\subsection{BRC~30}
This tightly curved rim is located in the Sh2-49
H{\sc ii} region (aka M16) at a distance of 2.2 kpc that contains the Ser OB1
association (Humphreys 1978). The embedded source has an
IRAS luminosity of $<$590 L$_\odot$, computed with upper limits at 60 and 100
$\mu$m.  According to Fig.~2, the FIR colors of BRC~30 are almost consistent
with those of UCH{\sc ii} regions, but its luminosity clearly indicates that
it is an object of intermediate mass or, more likely, low-mass.

BRC~30 has been observed 5 times and maser emission is present 
only in the last two runs. The emission line is peaked at 
V$\sim$23~km~s$^{-1}$, close to 
the systemic cloud velocity of $V_{\rm cl}$=24~km~s$^{-1}$ (Brand \& Blitz 
1993).  The instantaneous isotropic H$_2$O luminosity is $1.0\times
10^{-6}$~L$_\odot$ at a distance of 2.2 kpc. Figure~3 displays the position
of BRC~30 in the far-infrared vs. H$_2$O luminosity plot. The plot also shows
the distribution of the sample of 14 sources that have been monitored for up
to 13 years at Medicina and that cover a large range of far-infrared
luminosities (Brand et al. 2003).  Note that in this plot, the maser
luminosity has been computed using the spectra obtained during the whole
monitoring campaign and therefore provides a more significant estimate of the
true source output over the instantaneous values obtained in single
observations that due to the high variability of the maser emission (e.g.
Wouterloot et al. 1995, Claussen et al. 1996) also tends to be quite
variable.  BRC~30 falls below the straight line representing the best fit to
the data points of Brand et al., but as we have noted above, the IRAS
luminosity represents an upper limit.

\begin{figure}
\psfig{file=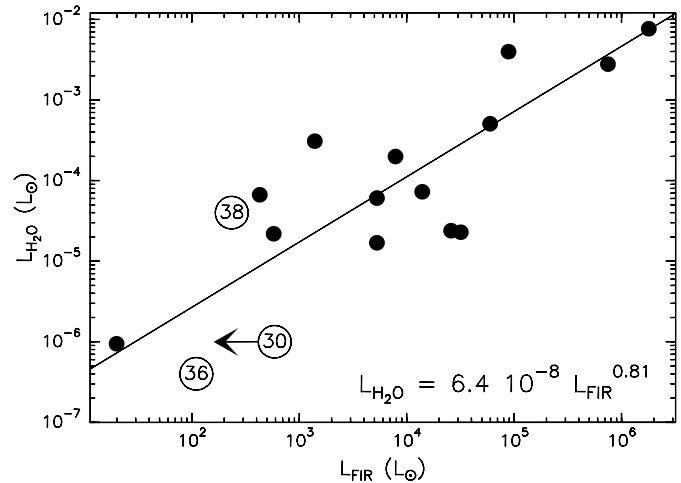,width=88mm,angle=-90}
\caption{Far-infrared vs. H$_2$O luminosity for the three BRC sources
detected at Medicina (empty circles with labels). The filled dots represent
the data points for the 14 sources monitored for up to 13 years (Valdettaro
et al. 2002). The solid line is a
least-squares fit to the maximum H$_2$O maser luminosity of these 14 sources 
derived by Brand et al. (2003).
\label{lum}}
\end{figure}

\subsection{BRC~36, IC~1396A, Elephant Trunk Nebula} 
BRC~36 is part of the IC~1396 complex, a nearby (d=750 pc) H{\sc ii} region
ionized by HD~206267 (O6.5V), the brightest star of the Trumpler 37 cluster
(Weikard et al.~1996).  IC~1396 contains many BRCs, 11 of which are listed in
SFO. BRC~36 hosts an embedded IRAS source with luminosity $L_{\rm FIR}=110
L_\odot$ and FIR colors typical of protostellar candidates or precursors to
UCH{\sc ii} regions (see Fig.~2).  Recently, high sensitivity observations
made with {\it Spitzer Space Telescope} in the NIR/MIR have revealed the
presence of a dozen embedded sources that contribute strongly at these and,
presumably, longer wavelengths (Reach et al. 2004).

BRC~36 has been observed five times and maser emission is present in three
occasions (see Fig.~1). The emission shows a single component at velocities
that vary between $-$6 and $-$10~km~s$^{-1}$, similar to the cloud velocity
of $-$8~km~s$^{-1}$ (Indrani et al. 1994). The integrated emission flux
displays a variation of a factor of 10 (see Table~2) between the three
observations (although it did not vary in the last two cases), a result not
too surprising considering that Wouterloot et al.  (1995) find variations of
up to 4 orders of magnitude and that in general low-luminosity sources tend
to have more variable emission than high-luminosity sources (Brand et al.
2003).  The location of BRC~36 in Fig.~3 below the correlation line can be
explained by the episodic emission and large flux variations. Clearly, a more
regular monitoring of this source would be desirable.

\subsection{BRC~38, IC~1396N}
BRC~38 also belongs to the IC~1396 complex and is located in the northern
part (thus called IC~1396N). The globule presents a striking cometary
structure and contains the source IRAS~21391$+$5802 with
a luminosity of 230 L$_\odot$ and a powerful molecular outflow (Codella et
al. 2001; SFO incorrectly give a value of 340 L$_\odot$). According to
Fig.~2, BRC~38 has the most extreme FIR colors of the entire BRC sample.

Water maser emission from this source has been observed by various
groups (Tofani et al. 1995, Slysh et al. 1999, Patel et al. 2000), following
the initial detection by Felli et al. (1992). These high angular resolution
observations, coupled with proper motion determinations, have established a 
solid physical relation between the water maser spots and the bipolar 
outflow from the intermediate luminosity IRAS source.

BRC~38 is part of the sample of about 40 sources in star forming regions that
have been monitored for more than 15 years with the Medicina radiotelescope
(see Valdettaro et al.  2002 for the initial results on 14 objects). As a
result, a total of 58 spectra of this source have been collected over such a
period and in all cases water maser emission was present. The spectrum shown
in Fig.~1 is just a typical one that highlights the salient features of the
emission with multiple lines over an extended velocity interval.  In this
respect, the H$_2$O maser in BRC~38 differs significantly from those found in
BRC~30 and BRC~36 even though the luminosity of the IRAS sources seems to be
similar.  However, as shown in Valdettaro et al. (2005b), the maser
emission properties of IC~1396N as derived from single-dish observations
are quite similar to those observed in other
sources of comparable luminosity.  The degree of variability of the integrated
emission is rather low, while variations in the velocity of the emission
features are substantial.  The position of BRC~38 in the plot of Fig.~3
reinforces the findings obtained by Brand et al. (2003) in other sources and
confirms the method for deriving reliable estimates of the total maser
luminosity using sources with frequent emission over long periods of time.

\section{Discussion}
Only three of the 44 BRCs in the northern hemisphere listed by SFO show
H$_2$O maser emission, a somewhat surprising result considering that these
globules are believed to be sites of intermediate- to high-mass star
formation (based on the luminosity of the associated IRAS sources, $10^2$ to
$10^4$ L$_\odot$) and that the maser detection rate is substantially high
towards massive SFRs (e.g., Palla et al. 1991; Palagi et al. 1993). However,
this negative result can be understood considering that the majority of the
IRAS sources have FIR colors outside the range suggested by Wood \&
Churchwell (1989) to represent massive protostars.  As shown in Fig.~2, many
BRCs occupy the region typical of ``Low" sources that are known to have very
low H$_2$O maser detection rates (e.g., Palla et al. 1993).

On the other hand, the lack of maser emission in 9 of the 10 sources with FIR
colors typical of UCH{\sc ii} regions is significant and not just due to the
limited sensitivity of the Medicina telescope. In fact, in the study of water
masers associated with candidate UCH{\sc ii} regions performed with the
Medicina antenna, Palla et al. (1991) obtained a detection rate of 26\%,
substantially higher than the present one (1/10=10\%). 
Moreover, even in the
case of the Nobeyama survey of low-luminosity YSOs conducted by Furuya et al.
(2001, 2003, a factor of 5--10 more sensitive than ours) all of the detected
masers showed emission well above the Medicina threshold ($\sim$1--3 Jy).

A possible explanation of the negative results on BRCs is that in most cases
the observed IRAS fluxes are not only due to the embedded YSOs, but also to
the emission from heated dust in the bright rim that surrounds the globule.
Since dense cores tend to be located near the head of the BRC, heating from
the PDR raises significantly the dust temperature and contributes to 
part of the FIR fluxes observed by IRAS.  A detailed radiative
transfer model for the case of BRC~38 supports this suggestion (Valdettaro
et al. 2005b).  Thus, the intrinsic IRAS luminosity of the embedded YSOs can
be substantially lower than the nominal value, and BRCs would then be sites
of low-mass star formation.  The extra heating from the warm dust in the rim
can also explain the lack of redshifted self-absorption in several BRCs
(De Vries e tal. 2002) which is considered a reliable signature of infalling
gas.

Our interpretation on the nature of the embedded sources finds direct support in the recent discovery of
groups of embedded sources within the Elephant Trunk Nebula (BRC~36) made
with {\it Spitzer} (Reach et al. 2004). In addition to the previously 
known H$\alpha$ emission line stars (LkH$\alpha$~349a and LkH$\alpha$~349c),
these observations have revealed the presence of about a dozen highly
embedded sources located near the dense core, close to the bright rim.
Although the individual bolometric luminosity of each source is still
unknown, from the shape of the SED extended to the mid-IR, it is clear that
they all contribute to the nominal IRAS fluxes.  Unfortunately, the poor
spatial resolution of the Medicina observations does not allow the
identification of the source of the H$_2$O maser emission, and
interferometric observations with the VLA towards this (and other) BRC would
be highly desirable in this respect.

Contrary to the ideas of induced intermediate- and high-mass star formation,
we argue that BRCs produce mostly {\it low-luminosity objects} for which the
frequency of occurrence of maser emission is low and highly episodic, as
shown by the results of our survey.  Furthermore, the embedded BRC sources
may be in a more advanced evolutionary phase.  The recent {\it Spitzer}
results on selected BRCs indicate the presence of several sources,
preferentially located in the vicinity of the bright rims. Although they have
been interpreted as protostars (e.g. Reach et al. 2004), their evolutionary
status is not clear.  Considering the higher H$_2$O maser frequency toward
Class 0 ($\sim$40\%) than Class I-II ($\sim$4\%-0\%, Furuya et al. 2003), the
negative finding of the Medicina survey seems to suggest that in general the
{\it Spitzer} embedded sources are not genuine protostars, but somewhat more
evolved objects. In such a case, the relatively short time of shock
propagation and compression ($\sim 10^4$ yr) is difficult to reconcile with
the idea of induced star formation within BRCs. In our view, the jury on this
mode of star formation is still out.

\begin{table*}
%{\sc H$_2$O Maser Observations of bright rimmed clouds} \\
\caption[]{H$_2$O Maser Observations of Bright Rimmed Clouds.\label{log}}
\begin{tabular}{llll}
\hline
\hline\noalign{\smallskip}
BRC      & IRAS   & Observation Date & r.m.s. \\
\#       & Source & (mm/yy)          &  (Jy) \\
\hline \hline
 1&23568$+$6706 & 01/91,01/93,01/94,12/00,05/04 & 1.2,1.0,0.8,1.9,1.3 \\
 2&00013$+$6817 & 01/93,12/00,05/01,04/04,05/04,04/05 & 1.1,1.2,4.1,0.9,0.7,3.7 \\
 3&00027$+$6700 & 01/93,12/00,05/01,05/04,04/05       & 1.1,1.2,4.2,0.7,1.8 \\
 4&00560$+$6037 & 04/92,12/00,05/01,05/04             & 1.1,1.2,5.5,1.2 \\
 5&02252$+$6120 & 11/89,01/99,12/00,05/01,05/04       & 2.7,0.6,1.2,4.2,1.1 \\
 6&02309$+$6034 & 01/93,12/00,05/01,05/04,12/04             & 1.0,1.2,4.4,1.3,0.5 \\
 7&02310$+$6133 & 01/93,12/00,05/01,05/04             & 1.0,1.2,4.8,1.5 \\
 8&02318$+$6106 & 01/93,12/00,05/01                   & 1.0,1.2,4.8 \\
 9&02326$+$6110 & 01/93,12/00,05/01,12/04             & 1.0,1.2,5.5,1.5 \\
10&02443$+$6012 & 01/93,12/00,05/01                   & 0.8,1.2,3.1 \\
11&02476$+$5950 & 01/93,12/00,05/01                   & 0.7,1.2,3.7 \\
12&02510$+$6023 & 01/93,12/00,05/01,06/04,12/04       & 0.8,1.2,3.5,1.7,0.5 \\
13&02570$+$6028 & 01/93,12/00,05/01,04/04,06/04,12/04 & 0.7,1.2,3.6,1.1,3.4,0.8 \\
14&02575$+$6017 & 03/89,11/89,04/90,07/90,04/91,12/00,05/01,04/04 
& 4.3,2.4,2.7,1.2,1.3,1.2,5.0,1.3\\
15&05202$+$3309 & 11/93,12/00,05/01,05/04       & 1.3,1.3,1.8,0.7,2.3,0.7 \\
16&05173$-$0555 & 02/90,04/98,12/00,05/01,05/04       & 0.9,2.3,1.6 \\ 
17&05286$+$1203 & 02/92,12/00,05/01,12/04             & 1.2,1.2,2.1,0.6 \\
18&05417$+$0907 & 02/90,12/00,05/01,03/05             & 1.0,1.3,2.1,0.9 \\
19&05320$-$0300 & 11/93,12/00,05/01,03/05             & 1.7,1.7,2.5,1.9 \\
20&05355$-$0146 & 04/90,12/00,05/01,03/05             & 2.0,1.7,2.5,1.6 \\
21&05371$-$0338 & 11/93,12/00,05/01,03/05             & 1.7,1.9,2.1,2.5 \\
22&05359$-$0515 & 11/93,12/00,05/01                   & 1.8,2.2,2.9 \\
23&06199$+$2311 & 12/91,12/00,05/01,06/04             & 0.6,1.2,1.6,0.8 \\
24&06322$+$0427 & 11/93,12/00,05/01,04/04,12/04,03/05 & 1.6,1.4,1.9,1.10.6,3.5 \\
25&06382$+$1017 & 05/91,12/00,05/01,04/04             & 1.9,1.4,1.9,1.1 \\
26&07014$-$1141 & 10/91,12/00,05/01,05/04,06/04,01/05 & 1.4,2.3,2.6,0.9,1.3,0.9 \\
27&07016$-$1118 & 10/91,01/93,12/00,05/01,05/04       & 1.5,1.2,2.6,2.7,1.0 \\
28&07023$-$1017 & 10/91,01/93,12/00,05/01,05/04,06/04 & 1.5,1.2,2.5,2.6,1.3,1.7 \\
29&07025$-$1204 & 05/91,10/91,12/00,05/01             & 2.2,1.6,3.6,2.9 \\
30&18159$-$1346 & 01/93,10/00,12/00,12/04,04/05       & 1.2,2.9,2.5,0.7,2.5 \\
31&20489$+$4410 & 01/92,12/00,05/01,05/04             & 0.9,1.4,2.8,0.9 \\
32&21308$+$5710 & 01/93,12/00,05/01,05/04             & 0.8,2.5,2.6,0.8 \\
33&21316$+$5716 & 01/93,12/00,05/01,05/04,09/04       & 0.8,2.6,1.8,0.7,0.9 \\
34&21320$+$5750 & 12/91,12/00,05/01,05/04,09/04       & 0.7,2.7,1.7,0.7,1.0 \\
35&21345$+$5818 & 01/93,12/00,05/01,05/04             & 0.8,2.9,1.8,0.7 \\
36&21346$+$5714 & 01/93,12/00,05/01,12/04,04/05       & 1.2,3.3,2.0,0.9,1.2 \\
37&21388$+$5622 & 10/90,05/91,12/00,05/01,09/04,12/04,04/05 & 1.9,1.4,1.9,1.9,3.6,0.7,0.5 \\
38&21391$+$5802 & total of 58 spectra    &         \\
39&21445$+$5712 & 02/90,04/90,12/00,05/01,10/02,12/02,04/03,11/03,04/04,01/05       & 1.3,1.6,1.7,3.0,2.4,1.0,1.3,1.3,0.8,1.0 \\
40&21446$+$5655 & 01/93,12/00,05/01,01/05,04/05       & 0.8,1.7,2.7,0.7,0.6 \\
41&21448$+$5704 & 01/93,12/00,06/05                   & 0.9,1.6,1.2 \\
42&21450$+$5658 & 01/93,12/00,06/05                   & 0.9,3.4,1.2 \\
43&22458$+$5746 & 01/93,12/00,06/05                   & 1.1,3.0,1.1 \\
44&22272$+$6358A& 01/89,02/90,01/94,03/99,12/00       & 4.8,1.1,0.9,0.6,2.6 \\
\hline 
\end{tabular}
\end{table*}

\begin{table*}
\caption[]{Properties of the H$_2$O masers in BRC 30 and 36.\label{prop}}
\begin{tabular}{c|c|c|c|c|c|c|c|c}
\hline
\hline
\multicolumn{9}{l}{}\\
%\multicolumn{9}{l}{}\\
\multicolumn{1}{c|}{\#} &
\multicolumn{1}{c|}{Date} &
   \multicolumn{1}{c|}{$\Delta V$} &
\multicolumn{1}{c|}{r.m.s.} &
   \multicolumn{1}{c|}{$V_{\rm min}$} &
\multicolumn{1}{c|}{$V_{\rm max}$} &
   \multicolumn{1}{c|}{$V_{\rm peak}$} &
\multicolumn{1}{c|}{$\int F_\nu {\rm d}V$} &
   \multicolumn{1}{c}{$L_{\rm H_2O}$} \\
%tabelhoofd \\ aan eind regel
\multicolumn{1}{c|}{} &
\multicolumn{1}{c|}{} &
\multicolumn{1}{c|}{(km s$^{-1}$)} &
\multicolumn{1}{c|}{(Jy)} &
\multicolumn{1}{c|}{(km s$^{-1}$)} &
\multicolumn{1}{c|}{(km s$^{-1}$)} &  
\multicolumn{1}{c|}{(km s$^{-1}$)} &
\multicolumn{1}{c|}{(Jy km s$^{-1}$)} & 
\multicolumn{1}{c}{(L$_\odot$)}  \\
\hline
  &  &  &  &  &  &  &  &   \\
30 &2004/12/18& 0.13  & 0.73 & 22.3    & 23.6   & 23.0    & 12.6 & 
$1.0\times 10^{-6}$\\
   &2005/04/14& 0.13  & 2.51 & 19.9    & 22.4   & 21.8    & 8.9  &
$7.7\times 10^{-7}$\\
  &  &  &  &  &  &  &  &   \\
36 &1993/01/28& 0.165  & 1.20 & $-$6.5  & $-$5.6 & $-$6.1  & 5.4  & 
$3.9\times 10^{-8}$\\
   &2004/12/18& 0.13  & 0.87 & $-$11.0 & $-$9.4 & $-$10.1 & 33.7 & 
$3.9\times 10^{-7}$\\
   &2005/04/14& 0.13  & 1.20 & $-$11.3 & $-$9.7 & $-$10.5 & 17.8 &
$2.1\times 10^{-7}$\\
\hline
\end{tabular}
\end{table*}

\begin{table*}
\caption[]{Properties of the BRCs with H$_2$O maser emission.\label{brc}}
\begin{tabular}{c|c|c|c|c|c}
\hline
\hline
\multicolumn{6}{l}{}\\
\multicolumn{1}{c|}{\#} &
\multicolumn{1}{c|}{SFR} &
\multicolumn{1}{c|}{IRAS} &
   \multicolumn{1}{c|}{d} &
\multicolumn{1}{c|}{$L_{\rm FIR}$} &
   \multicolumn{1}{c}{$V_{\rm cl}$} \\
%tabelhoofd \\ aan eind regel
\multicolumn{1}{c|}{} &
\multicolumn{1}{c|}{} &
\multicolumn{1}{c|}{} &
\multicolumn{1}{c|}{(kpc)} &
\multicolumn{1}{c|}{(L$_\odot$)}  &
\multicolumn{1}{c}{(km s$^{-1}$)} \\  
\hline
30 & Sh2-49  & 18159$-$1346 & 2.2  & $<$590 & $+$24 \\
  &  &  &  & &  \\
36 & IC~1396 & 21346$+$5714 & 0.75 & 110  & $-$8 \\
  &  &  &  & &  \\
38 & IC~1396 & 21391$+$5802 & 0.75 & 235  & $+$1 \\
\hline
\end{tabular}
\end{table*}

\bigskip
\noindent
{\it Acknowledgements.}~It is a pleasure to thank the staff of the Arcetri
radio astronomy group and of the Medicina station that have provided the usual 
professional and friendly
assistance with the observations during this survey.

\end{document}